\documentclass[
    ,final            
  ]
  {aipproc}

\layoutstyle{6x9}

\begin{document}

\title{Charmonium at CLEO}


\author{Kamal K. Seth}{
  address={Department of Physics and Astronomy, 
        Northwestern University, Evanston, IL, 60208, USA}
}

\begin{abstract}
At CLEO, charmonium spectroscopy is pursued both thorugh $e^+e^-$ annihilation data taken in the Upsilon region and more recently at $\psi(2S)$.  A nmber of first observations ($\eta_c'$, $h_c$, $\pi$ and $K$ form factors) have been made, and numerous high precision measurements have been made in radiative and hadronic decays of charmonium resonances.  A brief report of these contributions is presented.
\end{abstract}

\maketitle


\section{1. Introduction}

Charmonium was discovered in 1974.  During the next 10 years a great amount of discovery physics was done by the $e^+e^-$ colliders at SLAC, DESY and ORSAY.  In the 1990's the Fermilab $p\bar{p}$ experiments E760/E835 brought unprecedented mass resolution and precision to charmonium spectroscopy, and the BES detector also made numerous contributions at the Beijing $e^+e^-$ collider.  The BES II detector has since accumulated the world's largest data set with 59 million $J/\psi$ and 14 million $\psi(2S)$.  Unfortunately, the Fermilab experiments were limited because their detector did not have the capability to identify charged hadrons, and the BES detector had limited capability for photon detection.  In contrast, the CLEO detector, operating at the CESR $e^+e^-$ collider, is a state-of-the-art solenoidal detector with 93\% of $4\pi$ coverage.  It contains an excellent CsI electromagnetic calorimeter with $\sigma_E/E=1.5\%$ at 5 GeV (4\% at 100 MeV), excellent charged particle tracking with $\sigma_E/E\approx0.6\%$ at 1 GeV/$c$, and a RICH detector.  This allows for precision charmonium spectroscopy using $\sim15$ fb$^{-1}$ of $e^+e^-$ data in the bottomonium region ($\sqrt {s} = 9.46 - 11.30$ GeV), $\sim6$ pb$^{-1}$ of data taken at $\psi(2S)$ ($\sqrt{s}=3.686$ GeV), and $\sim20$ pb$^{-1}$ taken just belw $\psi(2S)$ ($\sqrt{s}=3.670$ GeV).

The list of measurements made at CLEO in the charmonium region is quite impressive.  They include discovery measurements and precision measurements as described in the following.  To put my presentation in perspective, the charmonium spectrum is shown in Fig. 1 (left).

\begin{figure}[!tb]
\includegraphics[width=2.8in]{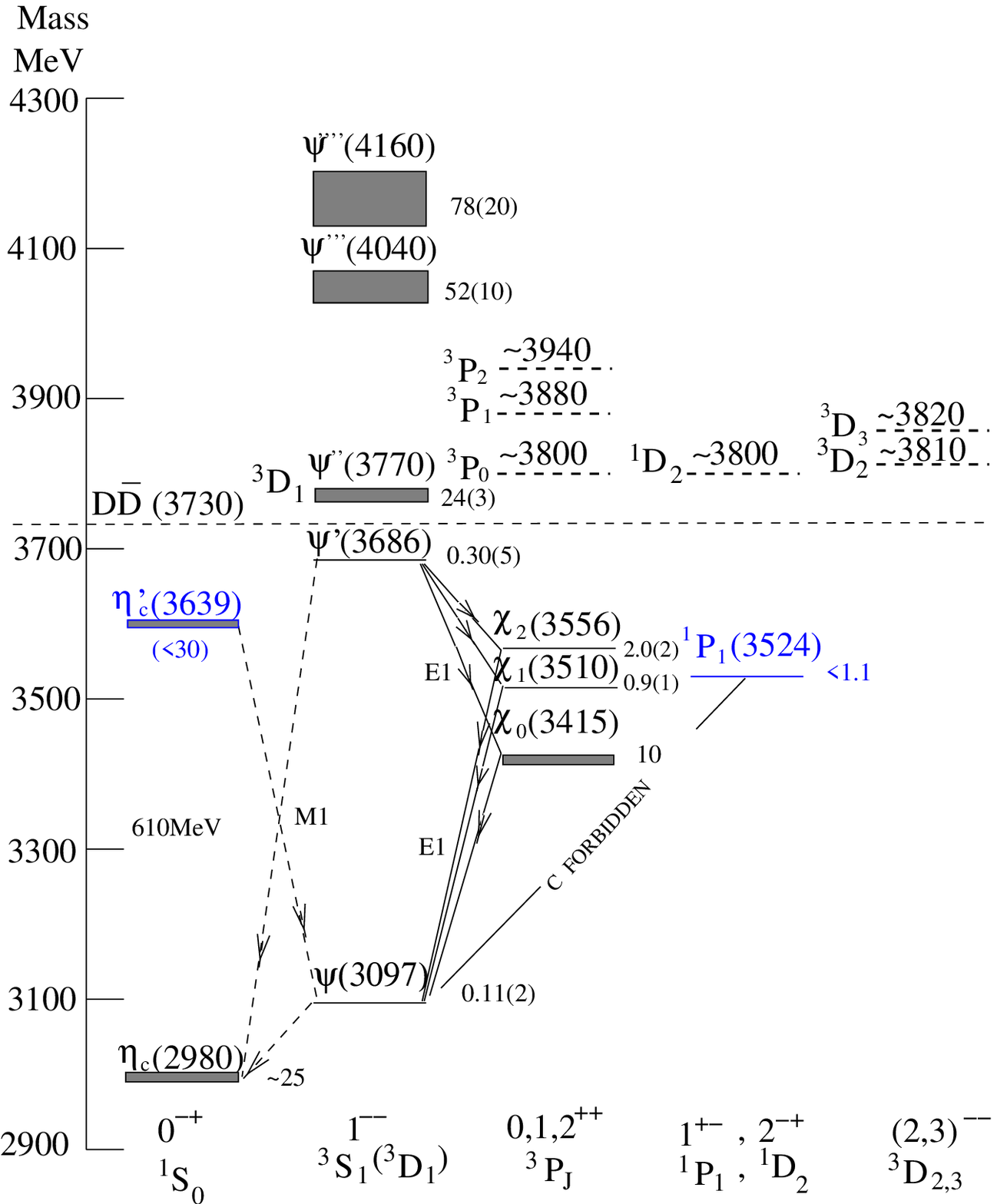}
\raisebox{0.2in}{\includegraphics[width=2.8in]{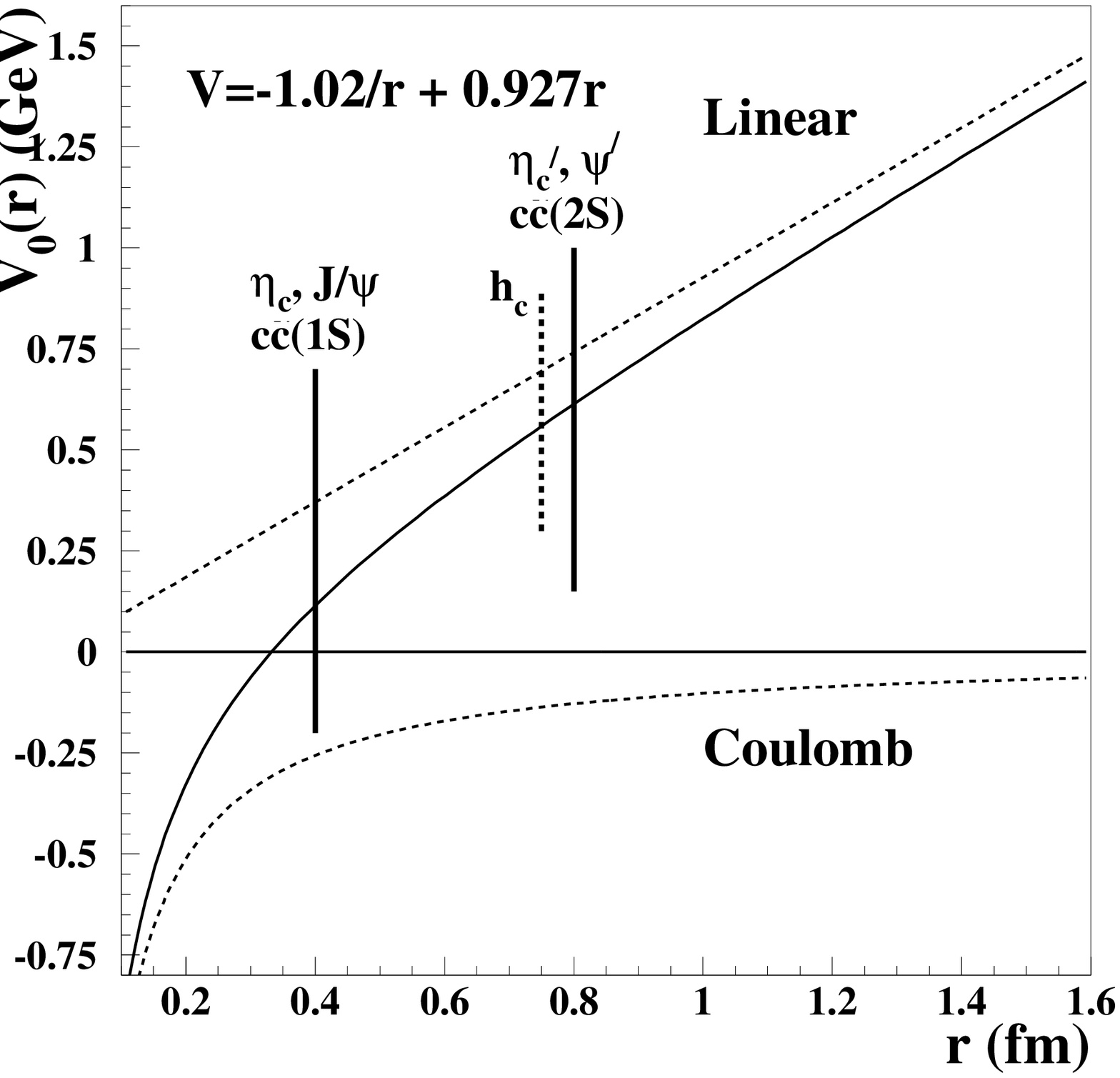}}
\caption{(left) Spectra of the states of Charmonium.  (right) Schematic of the QCD $q\bar{q}$ potential (solid line), and its Coulombic and confinement parts (dotted lines).  The approximate locations of the $1S$, $2S$ and $^1P_1$ states are shown.}
\end{figure}

\section{2. Discovery of $\eta_\MakeLowercase{c}'(2^1S_0)$ and $\MakeLowercase{h_c}(1^1P_1)$}

Spin singlet states are notoriously difficult to populate in $e^+e^-$ collisions.  For example, $\eta_c(1^1S_0)$ and $\eta_c'(2^1S_0)$ can only be reached by weak M1 radiative transitions from the directly produced vector states $J/\psi$ and $\psi(2S)$, and radiative transition to $h_c(1^1P_1)$ is forbidden by $C$--parity conservation.  However, identification of these states is crucial for understanding the spin--spin hyperfine interaction in the $q\bar{q}$ system.  The nature and development of the hyperfine interaction in going from $1S$ to $2S$ states, and in going from $S-$ to $P-$wave states, can only be detemined by identifying the spin--singlet partners of the well--established spin--triplet states, $J/\psi$, $\psi(2S)$, and $\chi_J(^3P_J)$.  As Fig. 1 (right) illustrates, the $2S$ and $1P$ states sample mainly the confinement part of the $q\bar{q}$ interaction, and determining their hyperfine splitting may be expected to shed light on the nature of the confinement interaction.
\vspace{12pt}

\noindent\textbf{2.1 The radial excitation of the charmonium ground state, $\eta_c'(2^1S_0)$}

\noindent Both $\eta_c(1^1S_0)$ and $J/\psi(^3S_1)$ states are well--established and measured.  The hyperfine splitting is known to be $\Delta M_{hf}(1S)=117\pm1$ MeV.  In contrast, while $\psi(2^3S_1)$, or $\psi'$, is well established with $M(\psi')=3686.11\pm0.03$ MeV, until now $\eta_c'(2^1S_0)$, or $\eta_c'$, has remained unidentified.  The Crystal Ball's \cite{cb-etacp} claim of the observation of $\eta_c'$ with $M(\eta_c')=3594\pm5$ MeV was never confirmed. Several subsequent attempts to find $\eta_c'$ were unsuccessful.  These included $p\bar{p}$ experiments E760 and E835 at Fermilab \cite{fnal-etacp}, two--photon fusion experiments at DELPHI and L3 at LEP, and a recent $\psi(2S)$ inclusive photon measurement at CLEO \cite{cleo-etacp}.  

\begin{figure}[!tb]
\begin{tabular}{cc}
\includegraphics[width=2.8in]{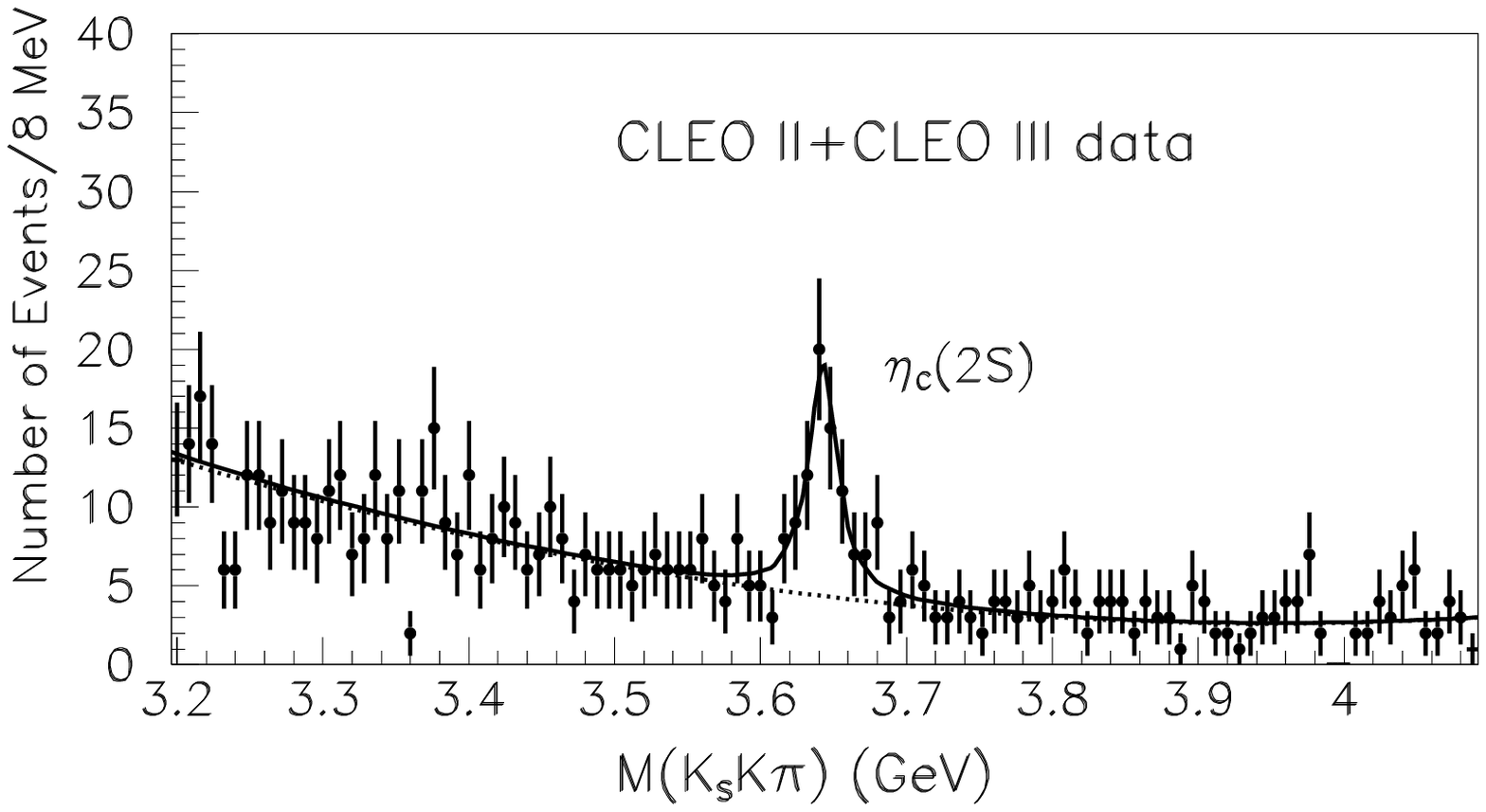}
&
\includegraphics[width=2.8in]{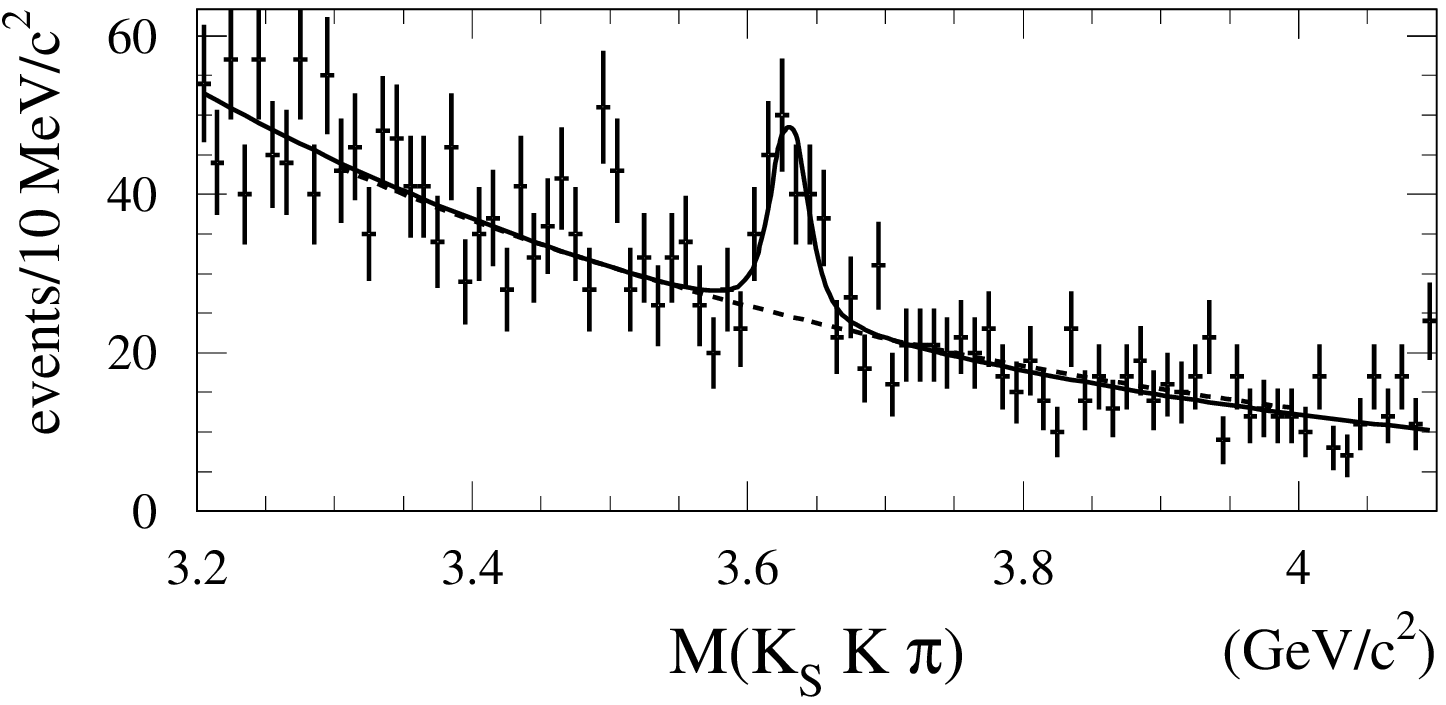}
\\
$M(\eta_c')=3642.9\pm3.4$ MeV \cite{cleo-2gam-etacp}
&
$M(\eta_c')=3630.8\pm3.5$ MeV \cite{babar-etacp}
\\
\end{tabular}
\caption{Observation of $\eta_c'$ in the reaction $\gamma\gamma\to K_SK\pi$ by CLEO (left) and BaBar (right).}
\end{figure}

The breakthrough came, of all the places, from the observation of $\eta_c'$ in $B$ decays by Belle \cite{belle-etacp}.  It was followed by its observation in $\gamma\gamma$ fusion by us at CLEO \cite{cleo-2gam-etacp}, and by BaBar \cite{babar-etacp} (see Fig. 2).  The present world average is $\left<M(\eta_c')\right>=3638.7\pm2.0$ MeV, which leads to the hyperfine splitting $\Delta M_{hf}(2S)=3686.1-3638.7=47.4\pm2.0$ MeV.  This is a factor $\sim2.5$ smaller than $\Delta M_{hf}(1S)$, and needs to be understood.  Is this due to configuration mixing or due to some special aspect of the hyperfine $\vec{s}_1\cdot\vec{s}_2$ interaction in the confinement region?  It is worth noting that the width of $\eta_c'$ is essentially unmeasured so far, and that it may shed light on the possible configuration mixing in $\eta_c'$.
\vspace{12pt}

\noindent\textbf{2.2 The singlet $P$ state, $h_c(1^1P_1)$, of charmonium}

\noindent If there ever was a more elusive state than $\eta_c'(2^1S_0)$, it is $h_c(1^1P_1)$, the singlet P-state of charmonium. Its identification is important to determine the hyperfine splitting of P-states.  And far greater precision in its mass determination than for $M(\eta_c')$ is required because it is predicted that $\Delta M_{hf}(1P) = M(\left<^3P_J\right>)-M(^1P_1)=0$ for the generally accepted scalar confining potential. 

Since the $h_c$, with $J^{PC}=1^{+-}$, cannot be populated in a radiative transition from the $1^{--}$ $\psi(2S)$ state, neither SLAC nor BES experiments have ever claimed to observe $h_c(^1P_1)$.  In 1992, the Fermilab $p\bar{p}$ experiment E760 claimed to have identified $h_c$ in the reaction $p\bar{p}\to h_c\to\pi^0 J/\psi$, but larger luminosity runs in 1996 and 2000 have failed to confirm this ``observation'' \cite{e835-hc}.

Now, we have made a firm (significance $>6\sigma$) observation of $h_c$ at CLEO \cite{cleo-hc}.  At CLEO-c data were taken at $\psi(2S)$, with 3.08 million $\psi(2S)$.  We have analyzed these data for the population of $h_c$ in the isospin forbidden reaction,
$$\psi(2S)\to\pi^0h_c\;,\; h_c\to\gamma\eta_c$$
Both inclusive and exclusive analyses were done, and an accurate determination of $h_c$ mass was made in recoils against $\pi^0$'s whose energy could be measured with precision (see Fig. 3).

Two independent \textbf{inclusive analyses}, different in details of event selection, were made.  The decay $h_c\to\gamma\eta_c$ was identfied by loosely constraining the photon energy in one analysis, and by loosely constraing the $\eta_c$ mass in the other analaysis.  Completely consistent results were obtained. In the \textbf{exclusive analysis}, instead of constraining $E_\gamma$ or $M(\eta_c)$, seven known decay channels with a total branching fraction of $\sim10\%$ were measured.  Once again, consistent results were obtained.

\begin{figure}[!tb]
\includegraphics[width=2.4in]{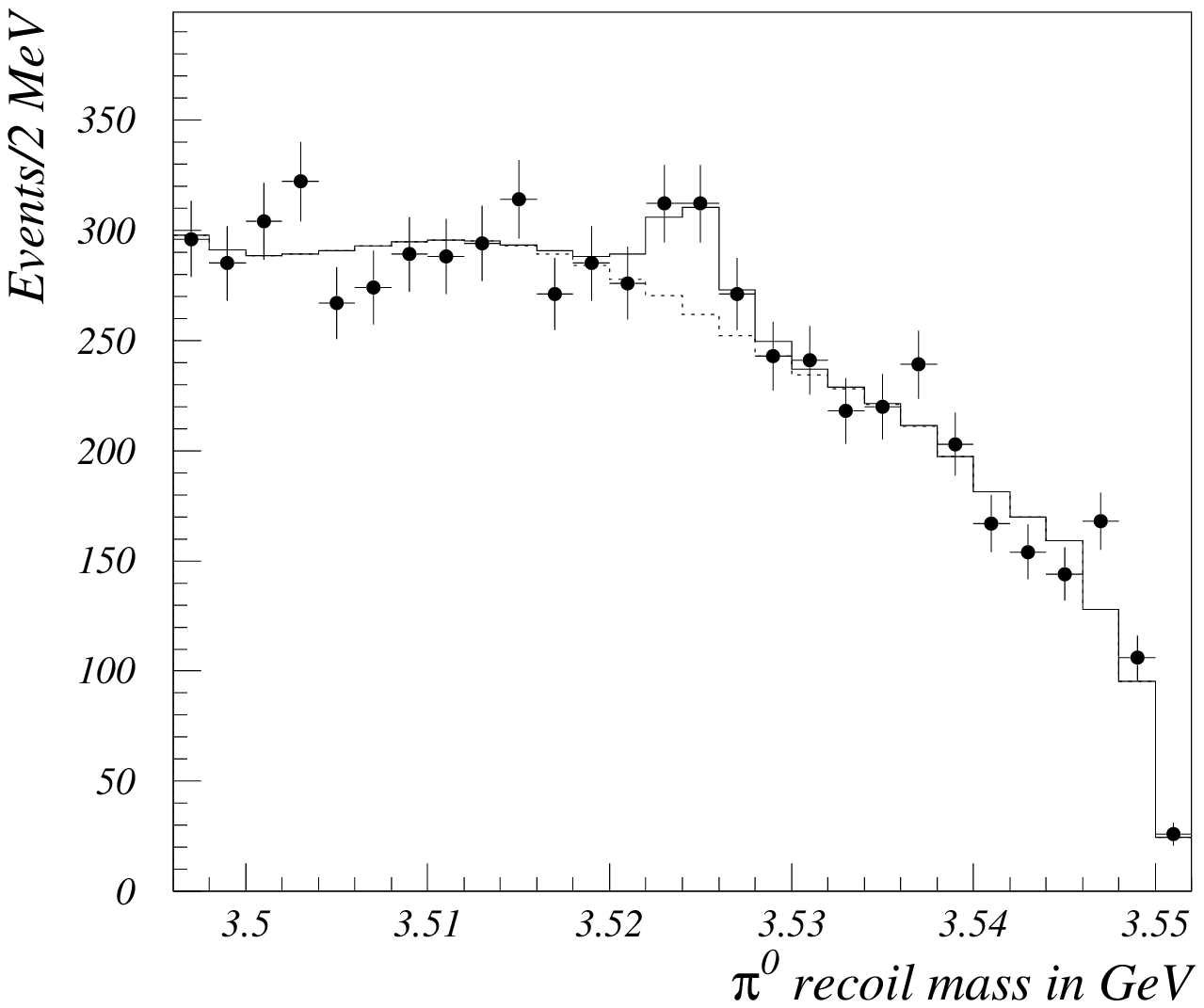}
\raisebox{1.8in}{\rotatebox{270}{\includegraphics[width=1.7in]{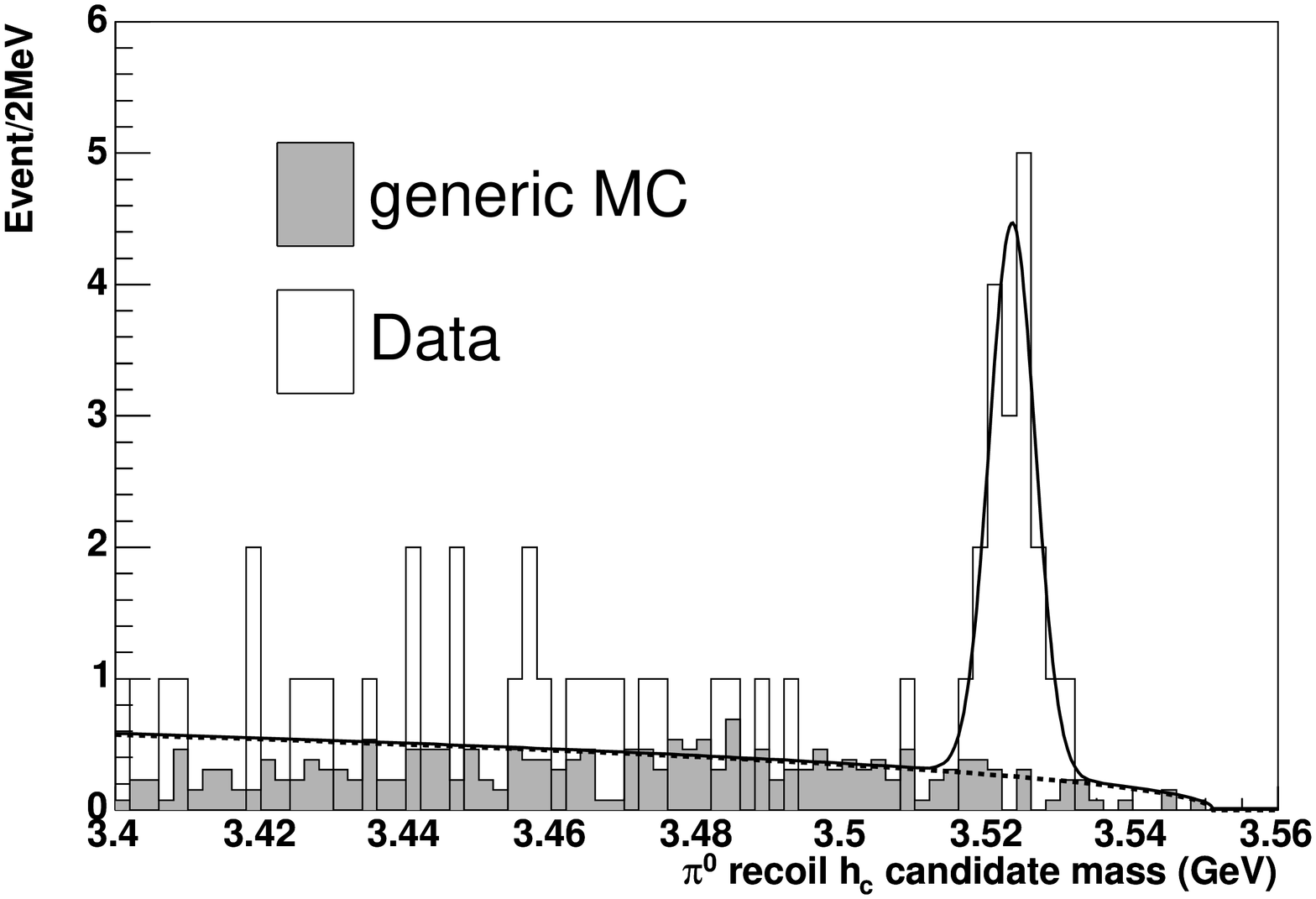}}}
\caption{Observation of $h_c(1^1P_1)$ in (left) inclusive analysis and (right) exclusive analysis at CLEO \cite{cleo-hc}.}
\end{figure}

The overall result obtained was $M(h_c) = 3524.4\pm0.6\pm0.4\;\mathrm{MeV}$, or
$$\Delta M_{hf}(1P)=\left<M(\chi_{cJ})\right>-M(h_c)=+1.0\pm0.6\pm0.4\;\mathrm{MeV}$$

Two results follow from our unambiguous observation of $h_c$.  The first is that $\Delta M_{hf}(1P)$ is not very different form the pQCD prediction of zero, as was feared by some theorists.  The second is that higher statistics data need to be taken at CLEO in order to arrive at a statistically significant result for $\Delta M_{hf}(1P)$.

We note that in a recent publication, E835 has claimed evidence for $h_c$ in the reaction $p\bar{p}\to h_c\to\gamma\eta_c$ in a combined analysis of data from their 1996 and 2000 runs.  However, only (7+6=) 13 counts were observed, and the significance of the observation was only $\sim3\sigma$ \cite{e835-hc}.

\section{3. Timelike Form Factors of $\pi$, $K$, and $\MakeLowercase{p}$}

Electromagnetic form factors of composite particles provide deep insight into their structure, and play an important role in determining the momenta above which perturbative QCD can be considered reliable, a question about which there has been great controversy.  

We have used the CLEO-c data for $e^+e^-$ collisions at $\sqrt{s}=3.671$ GeV with a total integrated luminosity of 20.7 pb$^{-1}$ to determine the timelike form factors of the kaon and proton at $|Q^2|=13.48$ GeV$^2$ \cite{cleo-ff}.  

The excellent charged particle detection capability of the CLEO detector has enabled us to determine
\begin{eqnarray*}
Q^2|F_\pi(13.48\;\mathrm{GeV}^2)| &=& 1.01\pm0.11(\mathrm{stat})\pm0.07(\mathrm{syst}),)\;\mathrm{GeV}^2,\\
Q^2|F_K(13.48\;\mathrm{GeV}^2)| &=& 0.85\pm0.05(\mathrm{stat})\pm0.02(\mathrm{syst})\;\mathrm{GeV}^2,\\
Q^4|G_M^p(13.48\;\mathrm{GeV}^2)|/\mu_p &=&0.91\pm0.13(\mathrm{stat})\pm0.06(\mathrm{syst})\;\mathrm{GeV}^4.
\end{eqnarray*}
The proton magnetic form factor result agrees with that measured in the reverse reaction $p\bar{p}\to e^+e^-$ at Fermilab as shown in Fig. 4 (left).  The pion and kaon form factor measurements are the first ever direct measurements at $|Q^2|>4.5$ GeV$^2$.

\begin{figure}[!tb]
\includegraphics[width=2.55in]{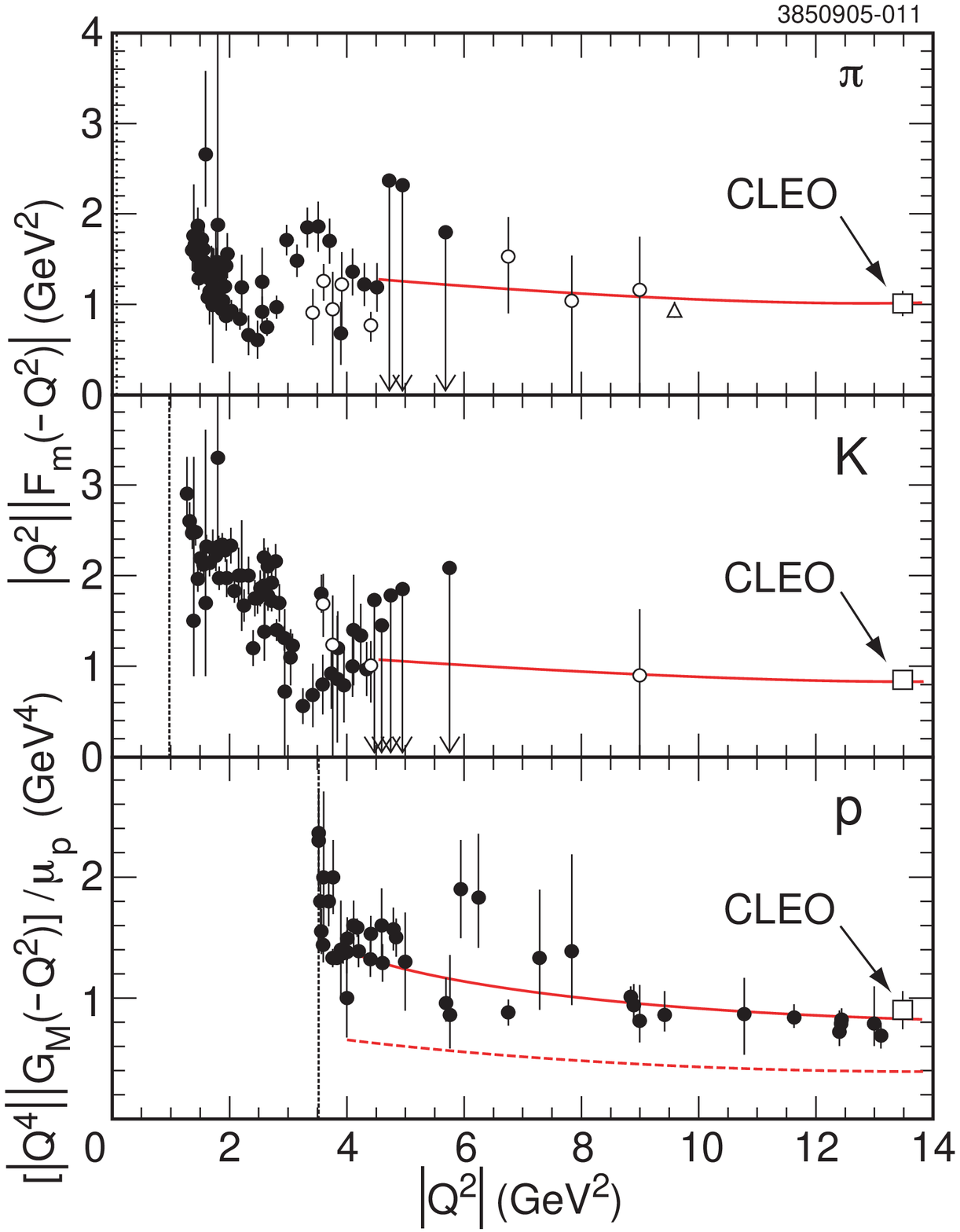}
\includegraphics[width=2.in]{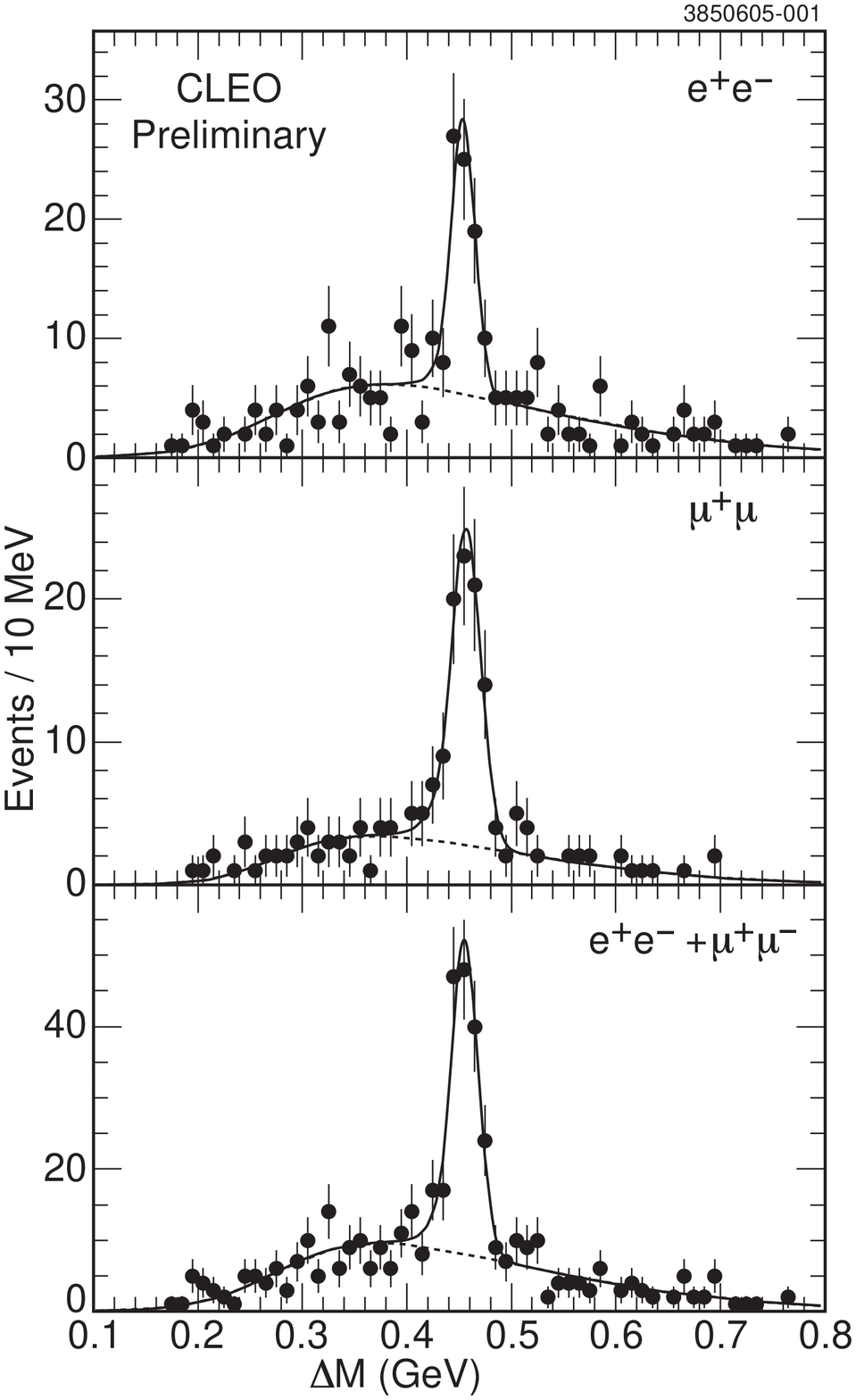}
\caption{(left) CLEO results (open squares) for the timelike form factors at $Q^2=13.48$ GeV$^2$ of the pion (top), kaon (middle), and proton (bottom) \cite{cleo-ff}.  Other results are from the literature.  Fig. 4 (right) Distributions of $\Delta M=M(\gamma l^+l^-)-M(l^+l^-)$ for the reaction $\gamma\gamma\to\chi_{c2}\to\gamma l^+l^-$ from CLEO \cite{cleo-chi2gam}.}
\end{figure}

\section{4. Radiative Decays of Charmonium}

In a recent publication, CLEO \cite{cleo-psirad} has reported the branching ratios, $\mathcal{B}(\psi(2S)\to\gamma\chi_{c0},\gamma\chi_{c1},\chi_{c2},\gamma\eta_c)$, from a careful analysis of the inclusive photon spectrum.  It is found that while $\mathcal{B}(\psi(2S)\to\gamma\chi_{c0},\gamma\chi_{c1},\gamma\eta_c)$ are in good agreement with earlier results \cite{pdg}, $\mathcal{B}(\psi(2S)\to\gamma\chi_{c2})$ is $\sim46\%$ larger.  This result, when combined with new CLEO \cite{cleo-psi-x} measurements of radiative cascades, $\psi(2S)\to\gamma\chi_{cJ}\to\gamma\gamma J/\psi$, leads to new results for the radiative transitions $\mathcal{B}(\chi_{cJ}\to\gamma J/\psi)$. These results are substantially different from the original Crystal Ball results which have long been in use.  In particular, it is found that $\mathcal{B}(\chi_{c2}\to\gamma J/\psi)=(19.9\pm1.7)\%$, as compared to the Crystal Ball result $\mathcal{B}(\chi_{c2}\to\gamma J/\psi)=(12.4\pm1.4)\%$.  This new CLEO result has resolved the long--standing discrepancy between the results for the two--photon width of $\chi_{c2}$ as measured at $e^+e^-$ colliders, via the production of $\chi_{c2}$ in two--photon fusion, and as measured in $p\bar{p}$ annihilation, via the decay of $\chi_{c2}$ into two photons.  Fig. 4 (right) shows the result of the new CLEO \cite{cleo-chi2gam} measurement of the reaction
$$\gamma\gamma\to\chi_{c2}\to\gamma J/\psi,\quad J/\psi\to(e^+e^-,\mu^+\mu^-)$$
using $\sim15$ fb$^{-1}$ of $e^+e^-$ collision data taken in the bottomonium region.  It is found that if the Belle two--photon fusion and E835 $p\bar{p}$ annihilation measurements are analyzed using the new CLEO value for $\mathcal{B}(\chi_{c2}\to\gamma J/\psi)$, their results come in complete agreement with the CLEO result, $\Gamma_{\gamma\gamma}(\chi_{c2})=559\pm81$ eV.  The CLEO result leads to $\Gamma(\chi_{c2}\to\gamma\gamma)/\Gamma(\chi_{c2}\to gg)=(3.65\pm0.45)\times10^{-4}$, and therefore $\alpha_S=0.29\pm0.01$ using pQCD with first order radiative corrections.

\section{5. The Saga of X(3872)}

In 2003 Belle \cite{bellex} announced the discovery of an unexpected narrow state, X(3872), in $B$--decays.  It was quickly confirmed by CDF, D\O, and BaBar.  The average of the masses measured by the four experiments is $M(\mathrm{X})=3871.5\pm0.4$ MeV.  Note that this is very close to $M(D^0)+M(\overline{D^{*0}})=3870.3\pm2.0$ MeV \cite{pdg}.  

The unique decay, the narrow width, and the closeness of its mass to $M(D^0D^{*0})$ have given rise to intense theoretical speculations about the nature of X(3872).  Is it a charmonium state ($1^{++}$, $2^{--}$, $3^{--}$), or a hybrid ($1^{++}$), or a glueball mixed with vector charmonium ($1^{--}$), or a unique `dimeson', a $D^0\overline{D^{*0}}$ ($1^{++}$, $0^{-+}$) molecule?  To sift through these speculations it is necessary to determine $J^{PC}$(X).

At CLEO \cite{cleox} we have analyzed $\sim15$ fb$^{-1}$ of $e^+e^-$ collision data taken in the bottomonium region for possible production of X(3872) in two--photon fusion and ISR, and have established the following 90\% confidence limits:\\
Two--photon fusion $(J^{PC}(\mathrm{X})=J^{P+})$: \qquad\qquad $(2J+1)\Gamma(\mathrm{X}\to e^+e^-)<0.65~\mathrm{eV}$,\\
ISR  $(J^{PC}(\mathrm{X})=1^{--})$: \qquad\qquad $\mathcal{B}(\mathrm{X}\to\pi^+\pi^- J/\psi)\times\Gamma(\mathrm{X}\to e^+e^-)<8.3~\mathrm{eV}$.\\

With BaBar's recent limit of $\mathcal{B}(\mathrm{X}\to\pi^+\pi^- J/\psi)>0.042$ \cite{babar-xdec}, the ISR result leads to $\Gamma(\mathrm{X}\to e^+e^-)<200~\mathrm{eV}$.  Recall that $\Gamma(\psi(3770))=260\pm40$ eV.

At CLEO an attempt is underway to measure the mass of $D^0$ with high precision, so that the proposed interpretation of X(3872) as a $D^0\overline{D^{*0}}$ molecule can be submitted to a critical test.

\section{6. $\psi(2S)$ Decays to Baryon--Antibaryon Pairs}

Very few decays of charmonia to baryon--antibaryon pairs have been measured.  Using 3.08 million $\psi(2S)$ decays, CLEO \cite{cleo-bary} has measured the decays $\psi(2S)\to B\bar{B}$  with $B\equiv p,\;\Lambda,\;\Sigma^+,\;\Sigma^0,\;\Xi^-,\;\Xi^0,\;(\Xi^{*0})\;\mathrm{and}\;(\Omega^-)$, many of them for the first time. A rather curious feature of these results is that the branching ratios for all $B\bar{B}$ pairs are nearly the same, $\sim\mathcal{B}(\psi(2S)\to B\bar{B})\approx 2.7\times10^{-4}$. This is surprising, considering that phase space is quite different for the different $B\bar{B}$ pairs.

\section{7. The $\rho-\pi$ Problem of $J/\psi(1S)$ and $\psi(2S)$}

A simple prediction of pQCD is that the ratio
$$Q\equiv \frac{\mathcal{B}(\psi(2S)\to\mathrm{hadrons})}{\mathcal{B}(J/\psi\to\mathrm{hadrons})} = \frac{\mathcal{B}(\psi(2S)\to\mathrm{leptons})}{\mathcal{B}(J/\psi\to\mathrm{leptons})} = 0.13\pm0.1.$$
BES has studied this over a long time by measuring many hadronic decays of $\psi(2S)$.  Recently CLEO \cite{cleo-had} has measured many more two--body and many body decays of $\psi(2S)$ using the data for $\sim3$ million $\psi(2S)$.  The results are that while $Q$ varies between 0.1\% and 10\% for two--body decays, it is generally larger, varying between 2\% and 10\% for most multibody decays.  For more details, see the talk by C. Z. Yuan in these proceedings.

\section{8. Precision Measurements of Hadronic Decays}

Spectroscopy does not progress only by discoveries.  It makes some of its best gains by precision.  Recently, several such gains have been made.  At CLEO \cite{cleo-psi-x}, isospin conservation has been confirmed at a $\pm2\%$ level in $\psi'$ decay to $\pi^+\pi^-J/\psi$ and $\pi^0\pi^0J/\psi$, and isospin violation has been observed at a $\pm0.4\%$ level in $\psi'$ decay to $\pi^0J/\psi$ and $\eta J/\psi$.  Also at CLEO \cite{cleo-psill} lepton universiality has been confirmed at a $\pm1\%$ level in $J/\psi$ decays to $e^+e^-$ and $\mu^+\mu^-$.

\section{9. Non--$D\overline{D}$ decays of $\psi(3770)$}

All that we know about $\psi(3770)$ is that it decays dominantly to $D\bar{D}$ ($\sim100\%$).  It is important to measure other possible decays of $\psi(3770)$.  $\psi(2S)$ decays to $\pi\pi J/\psi$ nearly 50\% of the time. What about $\psi(3770)$?  CLEO has invested 281 pb$^{-1}$ of luminosity, and obtained three new results.  It is determined that the branching fractions $\mathcal{B}(\psi(3770)\to\pi^+\pi^-J/\psi)=(0.214\pm0.033)\%$ and $\mathcal{B}(\psi(3770)\to\pi^0\pi^0J/\psi)=(0.097\pm0.040)\%$ \cite{3770-pion}.  The branching fraction $\mathcal{B}(\psi(3770)\to\gamma\chi_{c1})=(0.32\pm0.06\pm0.04)\%$ was measured, but only upper limits could be set for radiative decays to $\chi_{c2}$ and $\chi_{c0}$ \cite{3770-rad}.  No vector+psuedoscalar decays of $\psi(3770)$ were found (except for a hint of $\phi\eta$), and a search of 25 multibody final states also found none \cite{3770-had}.






\bibliographystyle{aipproc}   


\end{document}